\documentclass[%
 reprint,
nofootinbib,
 amsmath,amssymb,
 aps,
floatfix,
]{revtex4-1}

\usepackage{braket}
\usepackage{dsfont}
\usepackage{graphicx}% Include figure files
\graphicspath{{images/}}
\usepackage{dcolumn}% Align table columns on decimal point
\usepackage{bm}% bold math
\usepackage{mathtools}
\usepackage[thinc]{esdiff}
\usepackage{textcomp}
\usepackage{subfigure}
\usepackage{xcolor}
\usepackage{appendix}

\begin{document}

\title{Universality of $\mathds{Z}_3$ parafermions via edge mode interaction and quantum simulation of topological space evolution with Rydberg atoms}

\author{Asmae Benhemou$^1$}
\author{Toonyawat Angkhanawin$^2$}
\author{Charles S. Adams$^2$}
\author{Dan E. Browne$^1$}
\author{Jiannis K. Pachos$^3$}

\affiliation{
$^1$\textit{Department of Physics and Astronomy, University College London, London WC1E 6BT, United Kingdom}
}
\affiliation{$^2$\textit{Joint Quantum Centre (Durham-Newcastle), Department of Physics, Durham University, South Road, Durham, DH1 3LE, United Kingdom}}
\affiliation{$^3$ \textit{School of Physics and Astronomy, University of Leeds, Leeds LS2 9JT, United Kingdom}}

\date{\today}% It is always \today, today,

\begin{abstract}
Parafermions are $\mathds{Z}_{n}$ generalisations of Majorana quasiparticles, with fractional non-Abelian statistics. They can be used to encode topological qudits and perform Clifford operations by their braiding. Here we investigate the generation of quantum gates by allowing $\mathds{Z}_3$ parafermions to interact in order to achieve universality. In particular, we study the form of the non-topological gate that arises through direct short-range interaction of the parafermion edge modes in a $\mathds{Z}_3$ parafermion chain. We show that such an interaction gives rise to a dynamical phase gate on the encoded ground space, generating a non-Clifford gate which can be tuned to belong to even levels of the Clifford hierarchy. We illustrate how to access highly non-contextual states using this dynamical gate. Finally, we propose an experiment that simulates the braiding and dynamical evolutions of the $\mathds{Z}_3$ topological states with Rydberg atom technology. 

\end{abstract}
\maketitle
%In the framework of TQC, complementing the accessible set of Clifford gates with a non-Clifford operation. 

%\tableofcontents

\section{\label{sec:intro} Introduction}

Fault-tolerant quantum computing schemes were shown to exist using error-correcting techniques \cite{Kitaev20032, Knill_2005, 10.5555/874062} aimed at diminishing logical error rates by minimising the error rate on individual gates. In this context, physical systems that provide access to a set of exact elementary gates are advantageous. Topological quantum computation was introduced as a way to provide a computational framework for fault-tolerant quantum computation by Kitaev, Freedman and Preskill \cite{10.1007/3-540-49208-9_31, KitaevFreedmanPreskill}, which directly addresses the very low error rate requirement. The proposal is based on the use of anyons i.e. localised two-dimensional many body quantum systems that display exotic exchange statistics. While the braiding of Abelian anyons is characterised by an arbitrary phase factor, the statistics of non-Abelian anyon exchange are described by representations of the braid group. The non-Abelian character renders these objects useful for computation, which is carried out by creating pairs of anyons from the vacuum, inducing operations by adiabatically moving them around each other and fusing them, with the classical outcome defined by the resulting charge types and a very low estimated error rate. Quasi-particle modes emerging in condensed matter systems have been shown to carry (projective) non-Abelian statistics which can be identified with known anyon models. The most prominent example of such objects are Majorana zero-modes (MZMs) whose exchange is described by statistics of the Ising model \cite{Kitaev_2001, Kitaev_2006}, and which were found to appear in a two-dimensional electron gas in the Fractional Quantum Hall (FQH) regime \cite{MOORE1991362}, as one candidate for an experimental realisation. 

% Parafermions
Majorana fermions constitute the $\mathds{Z}_2$ case of the more general $\mathds{Z}_d$ parafermion model. The latter can be used to encode qudits, and provides a wider set of braiding evolutions, making it a more computationally powerful and attractive counterpart \cite{Fendley_2012}. Indeed, in contrast with the Majorana case \cite{Sarma_2015}, parafermions can provide a scalable entangling gate by topological operations. Much like MZMs, proposals to realise parafermions typically consist of exploiting the edge states of FQH systems \cite{Read1999, Vaezi_2014}, or generalizations of the Kitaev p-wave wire \cite{Kitaev_2001}.

% Need for universal gate 
Nevertheless, the braid group representation describing Ising and parafermionic statistics provides a reliable implementation of Clifford gates, does not extend to a universal quantum gate set and can therefore be efficiently simulated classically \cite{Bravyi_2005}. Proposals exist to remedy this drawback for Majorana qubits, by allowing for additional noisy non-topological operations which take the form of direct short-range edge mode interaction, i.e. a tunneling process. Such operations can give rise to the $\pi/8$-rotation that together with Clifford operations constitute a universal set \cite{Bravyi_2005}. Parafermions generalise the Majorana encoding to  topological qudits. For prime $d$ Clifford unitaries complemented by any arbitrary non-Clifford gate are sufficient for universal quantum computing (UQC) \cite{Anwar_2012}. Hence, the parafermion edge mode (PEM) interaction is expected to provide a noisy non-Clifford gate to be made fault-tolerant using magic state distillation (MSD) protocols. Such schemes have been extensively studied in particular for qudits of prime $d$.

Recent research has focused on quantum simulation with Rydberg atoms due to their versatility. They offer strong and controllable long-range interactions realized by selecting different Rydberg states and applying a wide range of optical fields \cite{Adams2019}. With the  development of experimental techniques improving the controllability of individual Rydberg atoms, such as optical tweezers, these systems represent an effective tool for simulating many-body physics of both coherent or dissipative, in- or out-of-equilibrium systems. This can be achieved by engineering the system Hamiltonian in order to simulate various spin systems and quantum phases of matter \cite{Browaeys2020, HansPeter2010, Harvard2021}. Lienhard et. al. and Verresen et. al.  \cite{Verresen2021, Lienhard2020} also suggested that geometric phases and topological effects can be probed with Rydberg atom-based quantum simulations. Additionally, Rydberg systems provide a way of encoding a qutrit by driving the Rydberg atom around three levels, using microwave lasers as described in Refs.~\cite{Adams2015, Adams2021}, which is of interest in the light of works such as Ref.~\cite{Gokhale2019}. 

In the following, we investigate which family of gates the interaction between the PEMs of a $\mathds{Z}_3$ parafermion chain gives rise to. Our main result concerns the adequacy of such gates for UQC with parafermions. We have chosen to focus on the three-dimensional qutrit space in this study since it has the benefit of prime dimension and computational tractability, though many of the features that we uncover are likely to be generic. Besides, we also suggest how to use a Rydberg atom to simulate topological evolution of the ground state of the parafermion chain Hamiltonian.

This paper is organised as follows. In Sec.~\ref{sec:bg}, we describe the $\mathds{Z}_3$ parafermion chain and its edge modes, then briefly describe computation with parafermions and the Clifford hierarchy. In Sec.~\ref{sec:PFU} we investigate the parafermion edge mode interaction and its action on the ground state space. In Sec.~\ref{sec:Mgate} we show that the addition of the dynamical gate available using the PEM interaction to the Clifford group, accessible through braiding operations, generates a gate-set dense in $SU(3)$. In Sec.~\ref{sec:Rydberg}, we discuss a potential physical implementation using a four-Rydberg level atomic system interacting with four microwave lasers, in order to simulate the direct parafermion interaction and two-parafermion braiding. Finally, our results are discussed in Sec.~\ref{sec:Discussion}.

\section{\label{sec:bg} Background}

\subsection{\label{subsec:chain} The parafermion chain}

In the following, we will consider a chain of $\mathds{Z}_3$ parafermions. In Ref.~\cite{Fendley_2012}, Fendley introduced a variation of the Kitaev chain expressed in terms of parafermion operators, whose Hamiltonian takes the general form 

\begin{equation}
    H = - \sum\limits_{j=1}^{L-1}J_j( \psi_j^{\dagger}\chi_{j+1}\alpha\bar{\omega} + \text{h.c} ) -\sum\limits_{j=1}^L f_j( \chi_j^{\dagger}\psi_j\hat{\alpha}\bar{\omega} + \text{h.c} )
    \label{eq:hamilt}
\end{equation}

\hspace{-4mm}where $L$ is the length of the chain, and at each site $j$ lie two parafermions $\chi_j$ and $\psi_j$. The $\omega = e^{\frac{2\pi i}{3}}$ factors ensure Hermiticity, and the couplings $f_j$ and $J_j$ are real and non-negative. The latter respectively characterise the flip and nearest-neighbour interaction Hamiltonian terms. The above Hamiltonian can be rewritten in terms of the chiral clock model, by re-expressing the parafermion operators as 
\begin{equation}
    \chi_j = \left(\prod_{k=1}^{j-1}\tau_k\right)\sigma_j \hspace{4mm}\text{and}\hspace{4mm}
    \psi_j = \omega\left(\prod_{k=1}^{j-1}\tau_k\right)\sigma_j\tau_j,
    \label{eq:pfops}
\end{equation}
where $\sigma$ and $\tau$ generalise the usual Pauli $\sigma^z$ and $\sigma^x$ matrices to a 3-dimensional space. For $j<k$, these operators follow the commutation relations
\begin{equation}
\chi_j\chi_k = \omega\chi_k\chi_j,\hspace{2mm}\psi_j\psi_k = \omega\psi_k\psi_j, \hspace{2mm}\chi_j\psi_k = \omega\psi_k\chi_j
\end{equation}
and one can verify that $\chi_j^3=\psi_j^3=1$ while each operator individually squares to its Hermitian conjugate. The three physical parameters of significance are the coupling ratio $f/J$, and two angles $\phi$, $\hat{\phi}$ which determine the parameters $\alpha$ and $\hat{\alpha}$. The integrable solution is given by the following
\begin{equation}
    \alpha_m = \frac{e^{i\phi(2m-n)}}{\text{sin}(\pi m/n)}\hspace{2mm}\text{and}\hspace{2mm}\hat{\alpha}_m = \frac{e^{i\hat{\phi}(2m-n)}}{\text{sin}(\pi m/n)}
    \label{eq:alphas}
\end{equation}
where $n = 3$ for the $\mathds{Z}_3$ chain, and the super-integrable case given by $\phi = \hat{\phi} = \frac{\pi}{2n}$, for which any values of $f/J$ are allowed \cite{Fendley_2012, Baxter_2006}. In the following we set $\alpha = \hat{\alpha} = e^{-\frac{i\pi}{6}}/\text{sin}(\pi/3)$ which lies at the mid-section between the ferromagnetic and anti-ferromagnetic phases of the model. Recent investigations of the parameter space of the system using DMRG tools offer detailed insight \cite{Zhuang_2015, Mahyaeh_2020}. 

\subsection{Parafermion edge modes}

When both time-reversal and spatial-parity symmetries are broken, parafermion zero-energy modes can emerge in the parafermion chain, localised at its edges. These are characteristic of topological order, and described by the left and right edge mode operators $\Psi_L$ and $\Psi_R$, which obey the relations  
 \begin{equation}
     [H,\Psi_{L}] = [H,\Psi_{R}] = 0,\hspace{3mm} \omega^P\Psi = \omega\Psi\omega^P
 \end{equation}
where $ \omega^P = \prod_{j=1}^{L}\tau_j^{\dagger} $ is the $\mathds{Z}_3$ symmetry generator. These properties respectively describe that these are zero energy modes, which map between $\mathds{Z}_3$ parity sectors, giving rise to the three-fold degeneracy of the energy spectrum of $H$. From Eq.~(\ref{eq:hamilt}) it is clear that exact PEMs exist in the limit of $f_i=0 \hspace{1mm} \forall \hspace{1mm} i$ such that $[H,\chi_1]_{f=0} = [H,\psi_L]_{f=0} = 0$ for a chain of length $L$. 

\begin{figure}[ht!]
    \centering
    \includegraphics[width=0.42\textwidth]{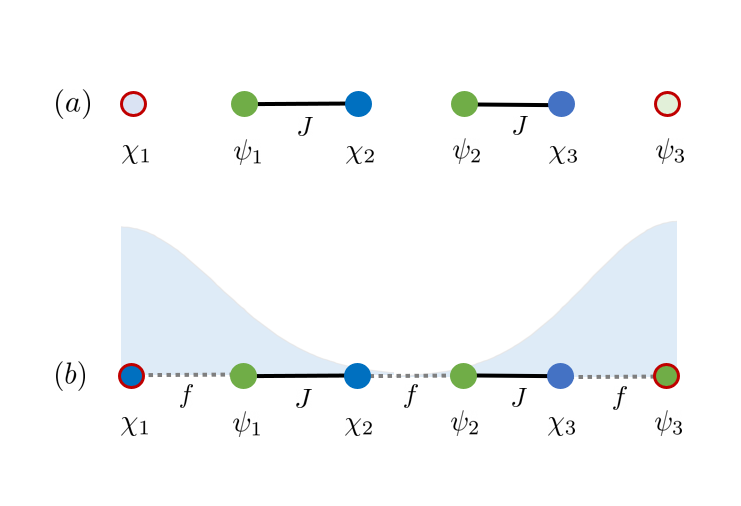}
    \caption{Schematic of the localised (a) and delocalised (b) edge modes, such that the latter are described by extended operators as introduced in Eq.~(\ref{eq:leftmode}), where the $\chi_i$ and $\psi_i$ operators represent left and right parafermions occupying each site of the chain, as per Eq.~\ref{eq:hamilt}. The shaded curves indicate their support on the bulk parafermions.} 
    \label{fig:pfchain}
\end{figure}

Since the system exhibits an energy gap, one expects its PEMs to remain approximate zero-modes for small enough $f/J$, with their support on the bulk of the chain exponentially suppressed in their distance to the bulk parafermions, as shown in Fig.~\ref{fig:pfchain} (b). The left edge mode operator of a chiral parafermion was constructed in Ref.~\cite{Fendley_2012} up to order $f/J$ using an iterative procedure which can be extended to all orders, and takes the form 
\begin{equation}
    \Psi_L = \chi_1 - 2ife^{-i\hat{\phi}}X + 2ife^{i\hat{\phi}}\chi_1^{\dagger}Y + ...
    \label{eq:leftmode}
\end{equation}
where
\begin{align}
  X &= \frac{1}{4J}\frac{1}{\text{sin}(3\phi)}(\psi_1 + e^{2i\phi}\chi_2 + e^{-2i\phi}\omega\psi_1^{\dagger}\chi_2^{\dagger} )
\end{align}
and $Y = -X^{\dagger}$. It is straightforward to show that the right edge mode can be derived in a similar fashion. One can see that the symmetric points $\phi = \frac{\pi}{6} \hspace{1mm}(\text{mod}\hspace{1mm}\frac{\pi}{3})$ are the most robust for the existence of edge modes. 

\subsection{\label{subsec:computing} Computing with parafermions}

\begin{figure}[ht]
    \centering
    \subfigure(a)\hspace{3mm}{\includegraphics[width=0.12\textwidth]{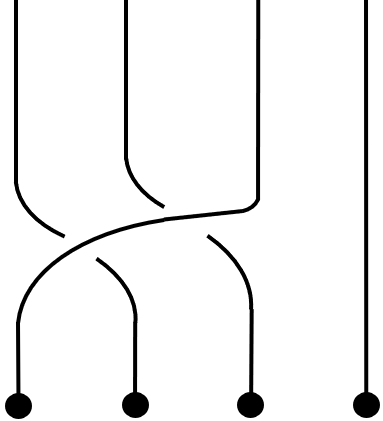}} 
    \hspace{10mm}
    \subfigure(b)\hspace{3mm}{\includegraphics[width=0.118\textwidth]{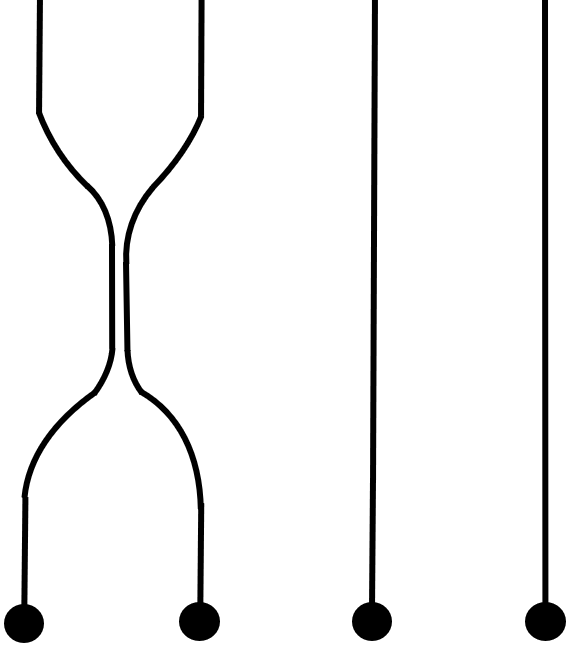}} 
    \caption{Schematic of single qutrit gates on a logical qutrit encoded in four parafermions denoted by the black dots. Their world-lines are depicted as time flows upwards. Sub-figure (a) represents a topological $S$ gate produced by parafermion braiding \cite{Hutter_2016}. In sub-figure (b) the closing distance between the world-lines of the first two parafermions represents the non-topological dynamical gate accessible by parafermion edge mode interaction.} 
    \label{fig:pfgates}
\end{figure}

Regardless of the underlying physical system which gives rise to parafermionic excitations, computation relies on the ability to adiabatically braid and fuse parafermions and distinguish between their fusion outcomes. Combining these capabilities realises a set of exact unitary operators and measurements. In Ref.~\cite{Hutter_2016}, Hutter and Loss derived the braiding operators acting on a logical qudit encoded using four $\mathds{Z}_n$ parafermions, and showed that for odd $n$ the braids $U_1$ and $U_1U_2U_1$ (up to global phases) are necessary and sufficient to generate the single qudit Clifford group, where in the $\mathds{Z}_3$ case specifically 
\begin{equation}
    U_i = \frac{1}{\sqrt{3}}\sum\limits_{m\in\mathds{Z}_3}c_m(\Lambda_i)^m ,
\end{equation}
where $\Lambda_i = \omega\chi_i\chi_{i+1}^{\dagger}$ are local parity operators for parafermions $\chi_i$ and $\chi_{i+1}$. In Fig.~\ref{fig:pfgates}(a) we show the $U_1U_2U_1$ braid which realises the topological $S$ gate. However, Clifford unitaries alone do not allow for universal quantum computing. In our study, we consider a similar approach to that in Ref.~\cite{Bravyi_2005}, namely allowing the parafermion modes to interact by bringing them close together so as to generate a dynamical non-Clifford unitary on the logical space. 
\vspace{5mm}

%Mention how this can be used between chains when qutrit encoded in a collection of parafermions.

\subsection{\label{subsec:CliffordH} Clifford hierarchy}

The theory behind quantum error correction and fault-tolerant computation relies upon access to gates from different levels of the Clifford hierarchy, defined by Gottesman and Chuang in Ref.~\cite{gottesman1999} as
\begin{equation}
    \mathcal{C}^{(k)} := \{ U|UPU^{\dagger} \in \mathcal{C}^{(k-1)}, \forall P \in \mathcal{P}_n \}
    \label{eq:hierarchy}
\end{equation}
where $\mathcal{P}_n:=\mathcal{P}^{\otimes n}$ in the $n$-qudit Pauli group, and $k$ indicates the level in the hierarchy \cite{1999gott}. With this definition, we can identify the Pauli and Clifford groups respectively as $\mathcal{C}^{(1)}$ and $\mathcal{C}^{(2)}$. Higher levels are of interest for any d-dimensional qudit system in order to achieve universality, and in particular any gate from the third level $\mathcal{C}^{(3)}$ supplementing the Clifford group generates a universal gate set. In the $d = 2$ case, the $T$-gate (i.e. $\pi/8$-phase gate) plays a special role as a natural choice \cite{Boykin2000}. While in general for $k \geq 3$ the gate-sets in the Clifford hierarchy do not form a group, the subset of diagonal operators $\mathcal{C}_d^{(k)} \subset \mathcal{C}^{(k)}$ does, making investigations for a T-gate qudit analogue tractable as shown in refs.~\cite{Howard_2012, Campbell_2012, Cui_2017}. In particular, Cui et. al. showed that a diagonal gate $U$ in any level of the Clifford hierarchy for qudits of dimension $d$ can be written as 

\begin{equation}
     U = \sum_{j\in\mathds{Z}_d}\text{exp}\left(2\pi i \sum_{m} \delta_m(j)/d^m\right)\ket{j}\bra{j},
\end{equation}

\hspace{-4mm}where $\delta_m(j)$ is a polynomial over $\mathds{Z}_d^m$, and the level of the Clifford hierarchy containing $U$ is given by the degree of $\delta_m(j)$ with the largest $m$ \cite{Cui_2017}. In the following, this definition is used to characterise the unitary operator we obtain from parafermion edge mode interaction.

\section{\label{sec:PFU} {A dynamical gate from parafermion interaction}}

The main object of this section is to understand how parafermion edge modes interact when the $f_i$ couplings in Eq.~(\ref{eq:hamilt}) are non-zero. This interaction ensures that one can induce a tunnelling process by transporting such PEMs within a sufficient distance of each other, running the interaction for a desired time interval and returning the anyons to their initial positions, whereby a dynamical gate is applied on a qutrit encoded in the degenerate ground subspace of the system \cite{Bravyi_2005, Sarma_2015}. 

\subsection{Decimation of the highest-energy term}
\label{subsection:decimation}

\begin{figure}[ht!]
    \centering
    \includegraphics[width=0.42\textwidth]{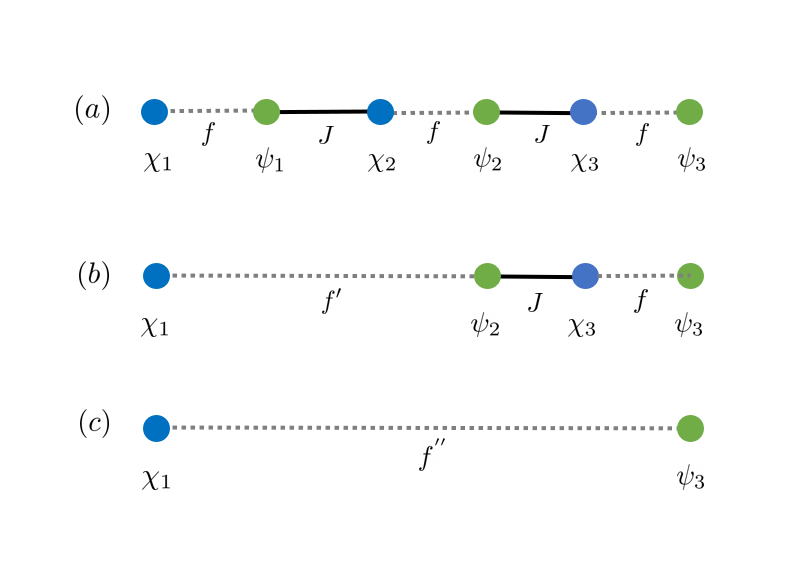}
    \caption{Schematic of a three-site parafermion chain with weak $f$ and strong $J$ couplings defined as per Eq.~(\ref{eq:hamilt}), where the $\chi_i$ and $\psi_i$ represent the left and right parafermion operators at each site of the chain (a). The chains in (b) and (c) represent the decimation of the terms connecting parafermions in the bulk upon applying the RG method, with a final weaker effective coupling $f''$.}
    \label{fig:pfchain2}
\end{figure}

The strongly interacting nature of parafermion systems makes their analytical study challenging, particularly the computation of their spectrum. A general approach is to use efficient DMRG techniques for numerical studies. In the following we use the real-space renormalisation group method applied to the transverse-field Ising model by Fisher in Ref.~\cite{Fisher1995}. This approach requires decimating the highest energy
term in the Hamiltonian and replacing it with effective longer range interactions. Specifically, in the case of the Ising chain this prescribes decimating a spin if the stronger interaction is a field $f_i$, or forming a ferromagnetic cluster if it is a bond $J_i$. We extend this scope to a parafermion chain where the $f_i$ on-site couplings are weak, such that the largest energy is the bond between chain sites as shown in Fig.~\ref{fig:pfchain2} (a) for a three-site chain. This process freezes the clock states at neighbouring sites together in a ferromagnetic cluster with an effective field $f'=\frac{f_if_{i+1}}{2J_{i,i+1}}$. This coupling is weaker than the individual $f_i$ and $f_{i+1}$ since the new interaction is a next-to-nearest neighbour one, which decimates the interaction between sites $i$ and $(i+1)$ (i.e. a parafermion pair). This process is illustrated in Fig.~\ref{fig:pfchain2} (a-c). Hence, the form of the Hamiltonian remains the same, apart from an overall constant shift in the spectrum, that can be neglected.

\subsection{\label{ssec:PFinteraction} {Effective Hamiltonian from PEM interaction}}

\begin{figure}
    \centering
    \includegraphics[width=0.47\textwidth]{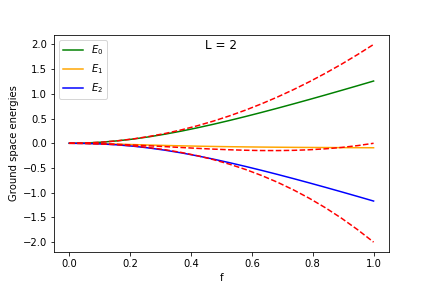}
    \caption{Three lowest energy eigenvalues for $H_2$ with $J_i=1$ $\forall i$ and varying $f$. The numerical values are indicated by the continuous lines. The perturbative approximations up to third order are indicated with red dotted lines, which re in agreement with the exact results for a wide range of $f$ values.}
    \label{fig:Lchainplot}
\end{figure}

We first consider a two-site $\mathds{Z}_3$-parafermion chain with a 
Hamiltonian $H_2$ describing the two-site version of Fig.~\ref{fig:pfchain2}(a), such that $H_2 = F_2 + V_2$ where
\begin{equation}
\begin{aligned}
    F_2 &= -\frac{2}{\sqrt{3}}e^{-i\phi}\bar{\omega}( J_1\psi_1^{\dagger}\chi_2) + \text{h.c.} \\
    V_2 &= -\frac{2}{\sqrt{3}} e^{-i\hat{\phi}}\bar{\omega}( f_1\chi_1^{\dagger}\psi_1 + f_2 \chi_2^{\dagger}\psi_2)  + \text{h.c.} 
\end{aligned}
    \label{eq:L2hamilt}
\end{equation} 
and we set the phase parameters to $\phi = \hat{\phi} = \frac{\pi}{6}$. To define the Hamiltonian we chose the representation
\begin{equation}
    \sigma = \begin{pmatrix} 1 & 0 & 0 \\
    0 & \omega & 0 \\
    0 & 0 & \omega^2
    \end{pmatrix}\hspace{3mm}\text{and}\hspace{3mm}\tau = \begin{pmatrix}
    0 & 0 & 1 \\ 
    1 & 0 & 0 \\
    0 & 1 & 0
    \end{pmatrix},
    \label{eq:sigmatau}
\end{equation}
to express the parafermion operators in Eq.~(\ref{eq:pfops}). The eigenspectrum of $F_2$ is three-fold degenerate, with eigenvalues $\{-2J_1, \hspace{1mm}0, \hspace{1mm}2J_1\}$ and three degenerate ground states $\{\ket{e_0}, \ket{e_1}, \ket{e_2}\}$. When $f_1$ and $f_2$ are non-zero, the edge modes of the parafermion chain are no longer exactly localised, as described by Eq.~(\ref{eq:leftmode}). A perturbative treatment of the effect of $F_2$ on the ground state manifold, in terms of $f_1/J_1$ and $f_2/J_2$, produces the effective coupling induced by an interaction between the edge parafermions $\chi_1$ and $\psi_2$ up to arbitrary order. The first order perturbation vanishes for our system, while the second order terms contribute to an effective Hamiltonian given by
\begin{equation}
    H^{(2)}_2 = -\frac{f_1^2+f_2^2}{J_1} + \frac{f_1f_2}{J_1}(\omega\chi_1\psi_2^{\dagger} + \bar{\omega}\psi_2\chi_1^{\dagger}).
    \label{eq:heffpf2}
\end{equation}
where we can see that the first term is a global energy shift which we neglect, and the second term characterises the interaction of the PZMs, which happens with a coupling $f' = \frac{f_1f_2}{J_1}$, and acts on the ground energy subspace with the following Hamiltonian
\begin{equation}
    H_2^{(2)} = \begin{pmatrix}
    0 & \omega & \bar{\omega} \\
    \bar{\omega} &  0 & \omega \\
    \omega & \bar{\omega} & 0
\end{pmatrix}.
\label{eq:Hint}
\end{equation}Following the prescription described in \ref{subsection:decimation}, the form of the second term in Eq.~(\ref{eq:heffpf2}) indicates that the decimation of the parafermion pair $(\psi_1, \chi_2)$ induces an $f'$ interaction between the edge parafermions. This result is similar to the Majorana case explicitly shown in Ref.~\cite{Vasudha2011} where the coupling between Majorana zero-modes takes an similar form for $L=2$. We iterate the above decimation procedure for a chain of arbitrary length $L$ by decimating bonds from left to right to derive the interaction between the edge modes of a chain of arbitrary size, as shown in Fig.~\ref{fig:pfchain2}. We find that the effective Hamiltonian up to second order in the energy perturbation is given by 
\begin{equation}
        H^{(2)}_{L} = -\frac{1}{J_{L-1}}\left( \frac{\prod\limits_{i=1}^{L-1}f_i^2}{\prod\limits_{i=1}^{L-2} J_i^2 } + f_L^2 \right) + \frac{\prod\limits_{i=1}^L f_i}{\prod\limits_{i=1}^{L-1} J_i}(\omega\chi_1\psi_L^{\dagger} + \bar{\omega}\psi_L\chi_1^{\dagger})
        \label{eq:Lchain}
\end{equation}
The PEM interaction term remains exactly $H_2^{(2)}$ in Eq.~(\ref{eq:Hint}), and has an eigenspectrum of $\{2,-1,-1\}$ which discriminates between the basis states $\{\ket{e_0}, \ket{e_1}, \ket{e_2}\}$ (up to a global $\mathds{Z}_3$ rotation). Similarly, the third order contribution to the effective Hamiltonian can be derived using the decimation procedure for a chain of $L$ sites, and takes the form
\begin{equation}
        H^{(3)}_{L} = -\frac{1}{\sqrt{3}}\left( f_L \prod\limits_{i=1}^{L-1} \frac{f_i^2}{J_i^2} + \frac{f_L^2}{J_{L-1}}\prod\limits_{i=1}^{L-1}\frac{f_i}{J_i}\right)(i\omega\chi_1\psi_L^{\dagger} + \text{h.c.})
        \label{eq:Lchain3}
\end{equation}
which acts non-trivially on the encoded ground space, with an eigenvalue spectrum of $\{0, 1,\hspace{1mm}-1\}$ up to a global phase given by the interaction coupling. The exact eigenvalues of the three lowest energy states of $H_2$ (for numerical convenience) are numerically plotted in Fig.~\ref{fig:Lchainplot} for $f_i = f$ $\forall$ $i$ and $J_i = 1$ $\forall$ $i$, where the global energy shift was deducted. The degeneracy is lifted by the splitting of the eigenstates when $f \neq 0$. The energy shifts obtained using the decimation prescription to study the effect of edge mode interaction are plotted up to third order in $f/J$, and agree with the exact spectrum in the perturbative, i.e. low $f/J$ regime.

\subsection{Asymmetric $\mathds{Z}_3$-parafermion chain}

In this subsection we report the same procedure as in \ref{ssec:PFinteraction} while allowing the parafermion chain to deviate from the super-integrable point of the chiral phase by setting $\phi = \pi/6$ and leaving $\hat{\phi}$ as a free parameter. In this case, the effective Hamiltonian for the  two-site chain up to second order in $f/J$ is given by
\begin{equation}
    H_2^{(2)}(\hat{\phi}) = -\frac{f_1^2 + f_2^2}{J_1} - \frac{f_1f_2}{J_1}(\bar{\omega}e^{2i\hat{\phi}}\chi_1\psi_2^{\dagger} + \omega e^{-2i\hat{\phi}}\psi_2\chi_1^{\dagger})
\end{equation}
\hspace{-4mm}such that the parafermion interaction takes the form 
\begin{equation}
    H_2^{(2)}(\hat{\phi}) = \begin{pmatrix}
    0 & e^{-2i\hat{\phi}} & e^{2i\hat{\phi}} \\
     e^{2i\hat{\phi}} & 0 & e^{-2i\hat{\phi}} \\
     e^{-2i\hat{\phi}} & e^{2i\hat{\phi}} & 0
    \end{pmatrix}
    \label{eq:phiInt}
\end{equation}
up to a global energy shift, which maintains the structure of $H_{2}^{(2)}$ in Eq.~(\ref{eq:heffpf2}), with an extra phase factor in the interaction term. The eigenvalues of $H_2^{(2)}(\hat{\phi})$ are given by: 
\begin{equation}
\begin{aligned}
    E_0 &= \hspace{3mm}2\hspace{1mm}\text{cos}(2\hat{\phi}) \\
    E_1 &= -\hspace{1mm}\text{cos}(2\hat{\phi}) +  \sqrt{3}\hspace{1mm}\text{sin}(2\hat{\phi}) \\
    E_2 &= -\hspace{1mm}\text{cos}(2\hat{\phi}) -  \sqrt{3}\hspace{1mm}\text{sin}(2\hat{\phi}),
    \label{eq:phiEvals}
\end{aligned}
\end{equation}
where setting $\hat{\phi} = \frac{\pi}{6}$ recovers the result in Eq.~(\ref{eq:Lchain}). This suggests that by allowing a margin of deviation closer to the ferromagnetic or anti-ferromagnetic phases of the system, one can modify the form of the interaction in Eq.~(\ref{eq:phiInt}). Specifically, setting the parameter $\hat{\phi} = \frac{\pi}{4}$ returns the eigen-structure of the third order interaction in the symmetric chain case in Eq.~(\ref{eq:Lchain3}) such that $H^{(2)}(\pi/4) = H^{(3)}$, which still exists in the chiral phase of the chain, i.e. the regime where PEMs subsist.  

\section{\label{sec:Mgate} Non-Clifford gate $\mathcal{U}$ from PEM interaction}

\subsection{\label{subsec:analytics} Dynamical gate}

The results in section \ref{ssec:PFinteraction} show that the interaction between the parafermions edge modes is described by a Hamiltonian which acts non-trivially on the encoded ground space, for which the leading term is shown in Eq.~(\ref{eq:Hint}).  Moreover, one can verify that this interaction (i.e.  $\omega\chi_1\psi_L^{\dagger} + \text{h.c.}$) commutes with the parity operator $\Lambda_1$ defined in subsection.~\ref{subsec:computing} and therefore preserves the computational subspace \cite{Hutter_2016}. Hence, bringing the parafermion edge modes closer together allows for a dynamical unitary operation given by the evolution of the interaction Hamiltonian
\begin{equation}
   \mathcal{U} \approx e^{-i\beta H_{int} t},
   \label{eq:Ugate} 
\end{equation}
where $\beta$ is a constant which depends on the effective coupling between the edge modes, namely up to second order $\prod\limits_{i=1}^L f_i/\prod\limits_{i=1}^{L-1} J_i$.

\subsection{\label{subsec:Uinhierarchy} $\mathcal{U}$ in the Clifford hierarchy}

In Ref.~\cite{Howard_2012} the authors derive explicit expressions for qudit (where $d$ is prime) analogues of the qubit $\frac{\pi}{8}$-phase gate such that these take a particular diagonal form, with the qutrit case given by
\begin{equation}
    U_v = U(v_0,v_1,...) = \sum\limits_{j=0}^{2} \zeta^{v_k}\ket{k}\bra{k} \hspace{2mm}(v_k \in \mathds{Z}_9),
\end{equation}
where $\zeta = e^{\frac{2\pi i}{9} }$, the indices $v_k$ are given by:  $v_0 = 0$ mod 9, $v_1=6z'+2\gamma'+3\epsilon'$ mod 9, $v_2=6z'+\gamma'+6\epsilon'$ mod 9 and $z',\gamma', \epsilon' \in \mathds{Z}_3$. We consider a three-dimensional state encoded in the ground state of such a chain. We then write the dynamical gate accessible on a qutrit from this interaction as the operator $\mathcal{U}$ which can be decomposed as 
\begin{equation}
    \mathcal{U} = H\hspace{1mm} \mathcal{U}_D H^{\dagger}  
    \label{eq:decomp}
\end{equation}
where $H$ is the qutrit Hadamard operator, Here $\mathcal{U}_D$ is the diagonal matrix
\begin{equation}
    \mathcal{U}_D \sim \begin{pmatrix}
    1 & 0 & 0 \\
    0 & 1 & 0 \\
    0 & 0 & e^{i\theta}
    \end{pmatrix}
\end{equation}
up to a global phase, where $\theta$ is specified by the interaction strength and duration in Eq.~(\ref{eq:Ugate}). Following Definition 3 in Ref.~\cite{Campbell_2012}, requiring that a qutrit $T$-gate be in $\mathcal{C}_3^{(3)}/\mathcal{C}_3^{(2)}$, i.e. belong in the third level of the Clifford hierarchy but not a Clifford gate, one requires that $TXT^{\dagger} \in \mathcal{C}^{(2)}$, where $X=\sum_{j\in\mathds{Z}_3}\ket{j + 1 \hspace{2mm}\text{mod}\hspace{1mm}3}\bra{j}$. Similarly,  $\mathcal{U}_DX\mathcal{U}_D^{\dagger} = XM$ where $M = \text{diag}\{1, e^{i\theta}, e^{-i\theta}\}$. In particular, for $\theta=\frac{2\pi}{3}$ and $\frac{2\pi}{9}$ respectively, $M = Z$ and $T$ where 
\begin{equation}
    Z=\sum\limits_{j\in\mathds{Z}_3}\omega^{j}\ket{j}\bra{j}
    \hspace{3mm}\text{and}\hspace{3mm}
    T=\sum\limits_{j\in\mathds{Z}_3}\zeta^{j^3}\ket{j}\bra{j}
\end{equation}
and $T$ is analogous to a qutrit $\frac{\pi}{8}$-phase gate, which indicates that $\mathcal{U}_D \in \mathcal{C}^{(4)}$. In general for a phase $\theta = \frac{2\pi}{3^m}$, $M = P_m(2)$ where we borrow the definition from Ref.~\cite{Cui_2017}:
\begin{equation}
    P_m(k) = \sum\limits_{j=0, k\neq j}^{p-1} \ket{j}\bra{j} + e^{\frac{2\pi i}{3^m}} \ket{k}\bra{k}
\end{equation}
where $P_m(k) \in \mathcal{D}_{m, p-1}$, the set of diagonal unitaries defined as 
\begin{equation}
    \mathcal{D}_{m,a} = \langle U_{m,b} \rangle_{b=1}^a . \{e^{i\phi}\} . \mathcal{D}_{m-1,p-1} 
\end{equation}
$$ \text{with}\hspace{2mm}     U_{m,b}:=\sum\limits_{j\in\mathds{Z}_p} \text{exp}(\frac{2\pi i}{p^m}j^b)\ket{j}\bra{j}.$$
where $\{e^{i\phi}\}$ accounts for all global phase shifts. Hence, $\mathcal{U}_D \in \mathcal{D}_{m,2}$. Cui et. al. also showed that for $m \in \mathds{N}$ and $1 \leq a \leq p-1$, $\mathcal{D}_{m,a} =  \mathcal{C}_d^{((p-1)(m-1) + a)}$. According to this result $\mathcal{D}_{m,2} = C_d^{(2m)}$, which signifies that for $\theta = \frac{2\pi}{3^m}$, the corresponding $\mathcal{U}_D$ belong to the $2m$-th levels of the Clifford hierarchy, i.e. $\mathcal{U}_D \in \mathcal{C}^{(2m)}$ so that for $m \geq 2$ $U_{D}$ is a non-Clifford operator. As stated in Eq.~(\ref{eq:decomp}), $\mathcal{U}$ is related to $\mathcal{U}_D$ by $H \in \mathcal{C}^{(2)}$. We therefore use Theorem 1 in Appendix \ref{app:huh} to show that for $\theta = \frac{2\pi}{3^m}$, $\mathcal{U} \in \mathcal{C}^{(2m)}$ as well. Setting $m=2$ gives rise to the gate $\mathcal{U}_{m=2}$ in $\mathcal{C}^{(4)}$ directly, from which implementing the qutrit $T$-gate in $\mathcal{C}^{(3)}$ is done with the combination $T = X^{\dagger}\mathcal{U}_{D}X\mathcal{U}_{D}^{\dagger}$. In order to obtain a set dense in $SU(d)$ one requires at least one non-Clifford element in the operational gate-set as shown in Ref.~\cite{Campbell_2012} for prime $d$ by combining results from Nebe, Rains and Sloane \cite{nebe_rains_a._2010}. The above illustrates how the parafermion interaction provides such gates, by choosing a parametrisation which creates a gate from low levels of the Clifford hierarchy, namely third and fourth. Additionally, we note that the form of the qutrit $T$-gate arises naturally from the eigenspectrum of the interaction Hamiltonian in Eq.~(\ref{eq:phiInt}) for $\hat{\phi}=\frac{\pi}{4}$. 

% Density plot 
\begin{figure}
    \centering
    \subfigure(a){\includegraphics[width=0.2\textwidth]{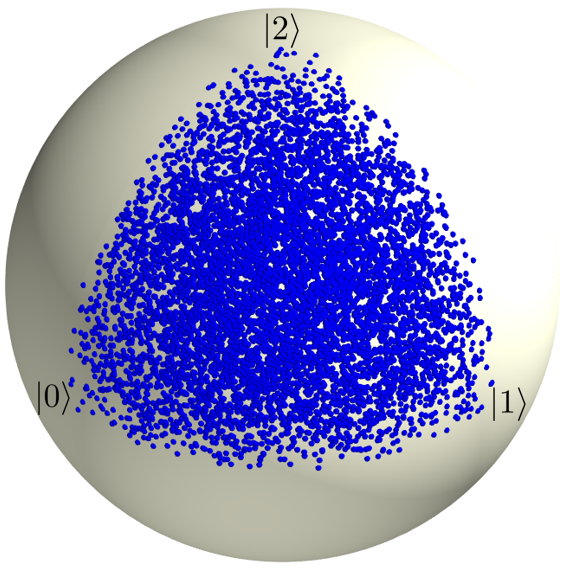}}
    \subfigure(b){\includegraphics[width=0.2\textwidth]{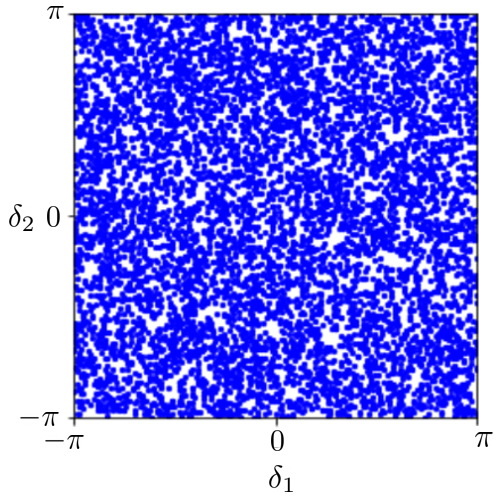}}
    \caption{Magnitude (a) and phase (b) plots  for 5000 sampled states computed by creating random words of length $50$ using $\mathcal{U}$ and Clifford operations, and applying them on the initial state $\ket{\psi}=\ket{0}$.} 
    \label{fig:sample1} 
\end{figure}

\begin{figure*}[ht!]
    \centering
    \subfigure(a){\includegraphics[width=0.25\textwidth]{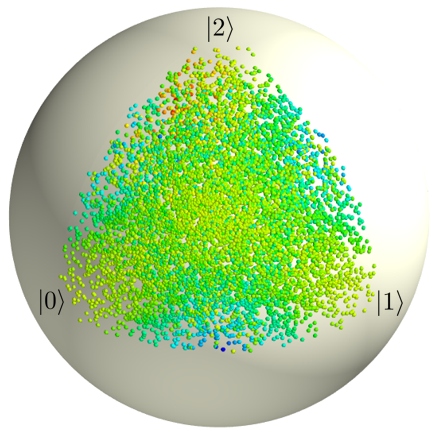}}
    \subfigure(b){\includegraphics[width=0.30\textwidth]{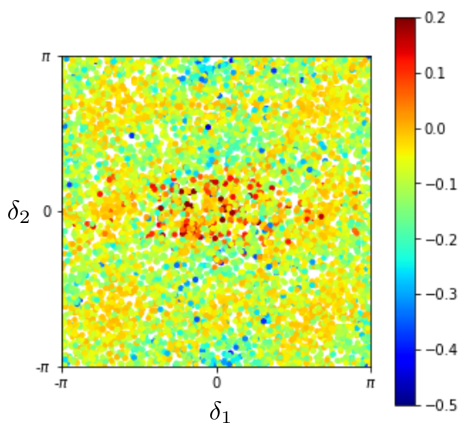}}
    \caption{Magnitude (a) and phase (b) plots of a dense set of $10 000$ sampled qutrit states using random words of length $50$ composed of the $\mathcal{U}$ and Clifford operations. The colour plot indicates the values of $\mathcal{M}$.}
    \label{fig:negstates}
\end{figure*} 

\begin{figure*}[ht!]
    \centering
    \subfigure(a){\includegraphics[width=0.24\textwidth]{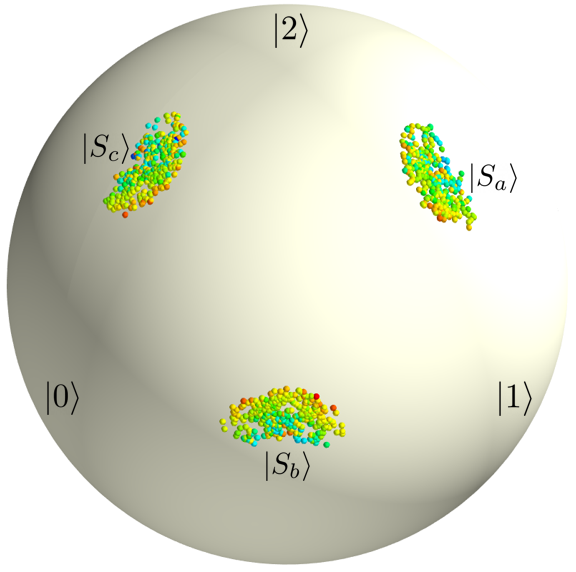}} 
    \subfigure(b){\includegraphics[width=0.305\textwidth]{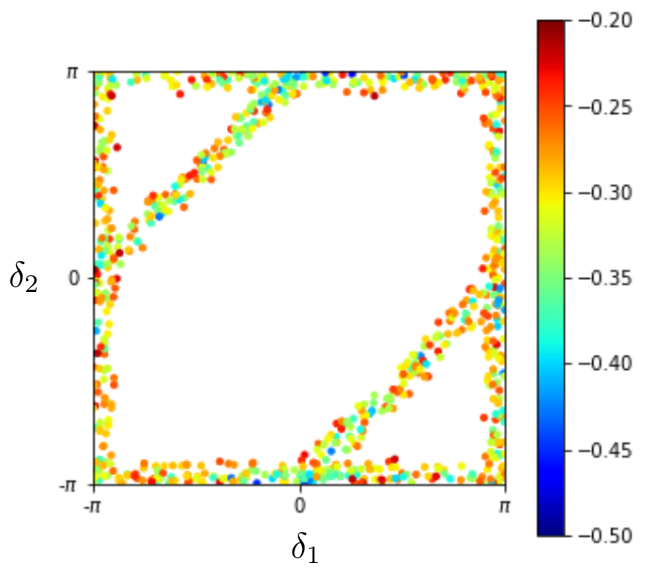}}
    \caption{Magnitude (a) and phase (b) plots of states close to the strange states $\ket{S_a}$, $\ket{S_b}$ and $\ket{S_c}$, within a trace distance of 0.2. The states were sampled using random words of length $50$ composed of the $\mathcal{U}$ and Clifford operations. The colour plot indicates the values of $\mathcal{M}$.} 
    \label{fig:distToS} 
\end{figure*}

\subsection{\label{subsec:brute} Universality with $\mathcal{U}$}

In order to study the universality of the parafermion computing scheme that encompasses PZM braiding as well as interaction, we created a series of words $W := U_1U_2...U_n$ such that $U_i$ is chosen randomly from the set $\{ X, S, H, \mathcal{U} \}$ where the first three elements are Clifford operators. In this representation $S$ and $H$ take the form
\begin{equation}
    S = \begin{pmatrix}
    1 & 0 & 0 \\
    0 & \omega & 0 \\
    0 & 0 & 1 
    \end{pmatrix},\hspace{2mm}
    H = \begin{pmatrix}
    1 & 1 & 1 \\
    1 & \omega & \bar{\omega} \\
    1 & \bar{\omega} & \omega
    \end{pmatrix}   
\end{equation}
generalising the qubit $\sqrt{Z}$ and Hadamard operators. The non-Clifford $\mathcal{U}$ of the set is given by Eq.~(\ref{eq:Ugate}) with $\theta=1$ for simplicity. We apply words $W$ to chosen initial qutrit states $\ket{\psi_{ini}}$ suitably obtained by initialising the parafermion system as described in Ref.~\cite{Hutter_2016}. The final states $\ket{\psi_{fin}} = W\ket{\psi_{ini}}$ can be written as 
$\ket{\psi_{fin}} = \alpha\ket{0} + e^{i\delta_1}\beta\ket{1} + e^{i\delta_2}\gamma\ket{2}$ where $\alpha,\beta,\gamma \in \mathds{R}$. We plot the resulting phases $(\delta_1,\delta_2)$ and magnitudes $(\alpha, \beta, \gamma)$ from the final states in Fig.~\ref{fig:sample1} (a) and (c) by sampling 5000 words of length $50$ and applying them to the initial state $\ket{\psi_{ini}} = \ket{0}$. The resulting parameter spaces are densely populated, which indicates the universality of the braiding Clifford operations supplemented with $\mathcal{U}$. 

\subsection{\label{subsec:contextuality} Non-contextual qutrit states using $\mathcal{U}$} 

With the consideration that the set $\{ X, S, H, \mathcal{U} \}$ is dense in $SU(3)$, and can be used to generate arbitrary elements in the qutrit state space using four parafermions, one can characterise such accessible states by their resourcefulness in the context of magic state distillation protocols (MSD). Indeed, MSD represents a powerful method to distill reliable quantum states from multiple noisy counterparts, which proves useful in our context since contrary to the Clifford operations, the dynamical gate $\mathcal{U}$ is non-topological. Moreover, MSD can also be used to obtain useful non-stabiliser states granted one accesses a state in the appropriate distillable region to the former. In Ref.~\cite{veitch2012negative}, Veitch et. al. showed that all qutrit states with positive Wigner function are undistillable, and the positive region was charted out in Ref.~\cite{Anwar_2012}. In particular, Howard et. al. argued in Ref.~\cite{Howard_2014} that contextuality supplies the magic for quantum computation, and introduced the following contextuality measure
\begin{equation}
    \mathcal{M} \equiv \hspace{1mm} \underset{\textbf{r}}{\text{max}} \hspace{1mm}\text{Tr}[A^{\text{\textbf{r}}}\rho] \stackrel{\text{NCHV}}{\leq} 0
    \label{eq:howardM}
\end{equation}
where $\rho$ is an arbitrary qutrit state and the $A^{\textbf{r}}$ are projectors onto eigenstates of the qutrit stabilizer operators given by $ A^{x,z} = D^{x,z}A^{0,0}D^{x,z^{\dagger}} $ where $x,z \in \mathds{Z}_3$ and $A^{0,0} = \frac{1}{3}\sum_{x,z}D^{x,z}$ using $ D^{x,z} = \omega^{2xz}X^xZ^z $, where $X$ and $Z$ are the qutrit Pauli operators. We sampled a large set of qutrit states starting from a random choice of basis states, and plot the negative Wigner states, i.e. violating Eq.~(\ref{eq:howardM}) indicating their contextuality measure in Fig.~\ref{fig:negstates}. 
In Fig.~\ref{fig:negstates}, we see an overlap of points with different $\mathcal{M}$. This is clearly due to the fact that the phase and magnitude projections introduced in the figures above do not constitute a direct map of the state space, though are useful to witness the coverage density of the Hilbert space accessible with our parafermion universal gate set. Indeed, we use this representation as a witness that our gate set provides access to a dense set of negative, and therefore potentially distillable states. In fact states which considerably violate Eq.~(\ref{eq:howardM}) can be obtained from short gate combinations (eg. $\mathcal{U}(t_1)H\mathcal{U}(t_2)\ket{0}$ given optimal evolution times $t_1$ and $t_2$). We note from Fig.~\ref{fig:negstates} that the majority of the 10000 sample states created using the full gate-set fulfill the condition $\mathcal{M} < 0$.

There exists a class of states which maximally violates Eq.~(\ref{eq:howardM}), namely the ``strange'' states defined by having one negative Wigner function entry of value $-1/3$ \cite{Veitch_2014} such as $ \ket{S_a} = \frac{\ket{1}-\ket{2}}{\sqrt{2}}$ \cite{Jain_2020}. These are the eigenstates of the qutrit Fourier transform,
and are located at the mid-sections between basis states on the magnitude plot as indicated in Fig.~\ref{fig:distToS}, with $\mathcal{M} = -0.5$. Finding a combination of gates to generate $\ket{S}$ exactly is a non-trivial problem. However, we show in Fig.~\ref{fig:sample1} that using $\mathcal{U}$ and Clifford gate combinations one can access states close to the strange states. We indicate in Fig.~\ref{fig:distToS} the points of maximal $\mathcal{M}$ by labels $\ket{S_{a,b,c}}$, and characterise this using the trace distance, i.e. $\mathcal{D} = \text{Tr}(\rho,\sigma) = \frac{1}{2}||\rho - \sigma ||_1$ where $\rho$ and $\sigma$ are the density matrices of the sampled states and $\ket{S_{a,b,c}}$ respectively. In Fig.~\ref{fig:distToS} states with $\text{Tr}(\rho,\sigma) < 0.2$ are plotted to indicate the regions of interest, and Table.~\ref{table:dist2Strange} summarises the trace distances and $\mathcal{M}$ for the three points closest to the strange states in Fig.~\ref{fig:distToS}. We note that both points admit a close to maximal violation of Eq.~(\ref{eq:howardM}).

\begin{table} [h!]

\begin{center}
\begin{tabular}{||c c c c||}
 \hline
  & $\ket{S_a}$ & $\ket{S_b}$ & $\ket{S_c}$ \\ [0.5ex] 
 \hline\hline
 $\mathcal{D}$ & 0.059 & 0.057 & 0.035 \\ 
 \hline
 $\mathcal{M}$ & -0.436 & -0.422 & -0.469 \\
 \hline
\end{tabular}
\end{center}
\caption{Trace distances $\mathcal{D}$ and non-contextuality measure $\mathcal{M}$ of the three sampled states in Fig.~\ref{fig:distToS} closest to $\ket{S_a}$, $\ket{S_b}$ and $\ket{S_c}$.} 
\label{table:dist2Strange}
\end{table}

Hence, using the Clifford gates supplemented with $\mathcal{U}$ one can possibly reach states arbitrarily close to the strange states, allowing for large enough gate sequences, and time evolutions in Eq.~(\ref{eq:Ugate}).  Recent results in Ref.~\cite{Prakash2020} show how one can distill strange states using an MSD protocol (Ternary Golay Code \cite{PLESS1968215}) with particularly high threshold. This showcases the advantage of accessign such states. 

\section{\label{sec:Rydberg} Quantum simulation with Rydberg atoms} 
\begin{figure}[ht]
	\centering
	\includegraphics[width=3.5cm]{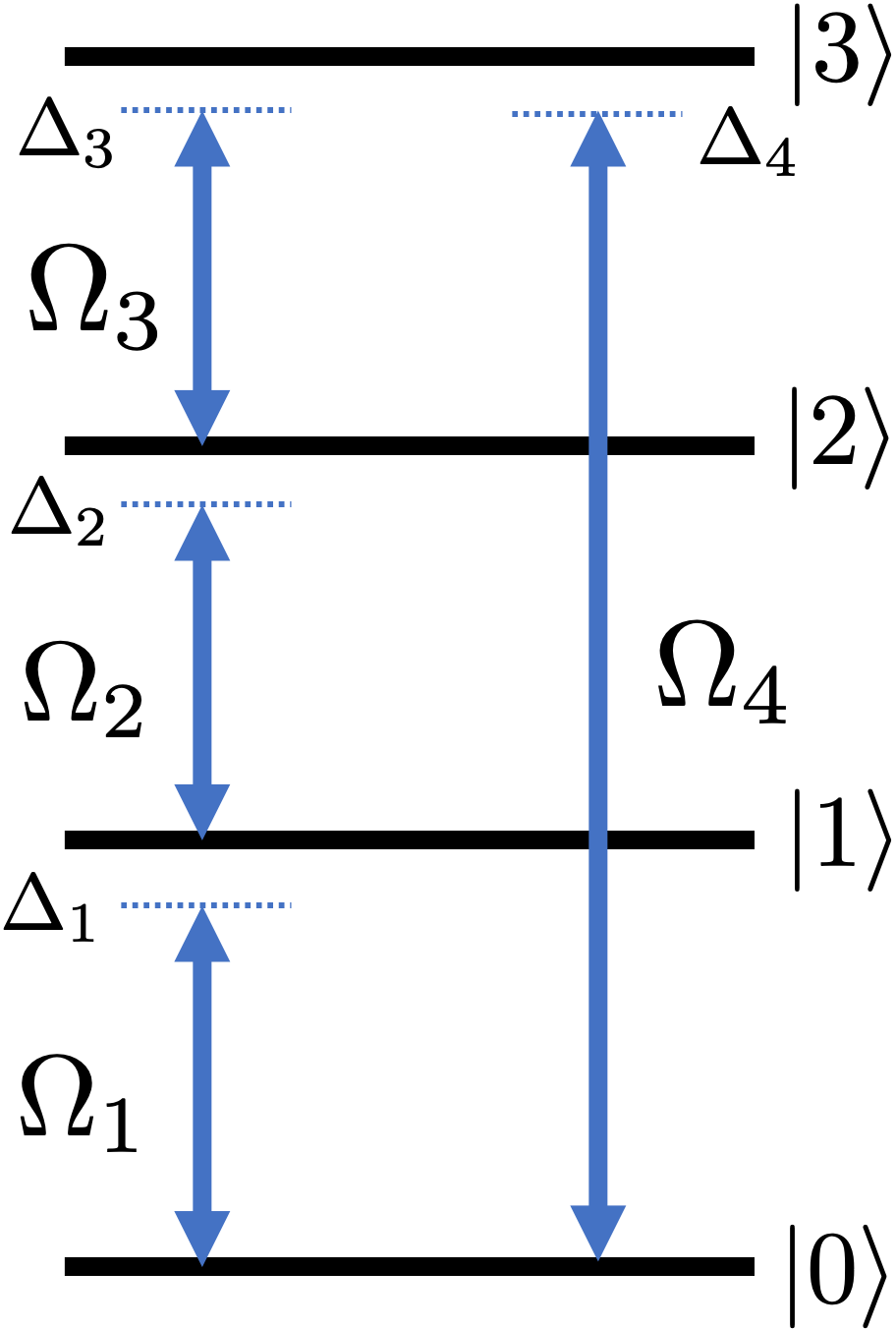}
	\caption{The interaction between a four-Rydberg level atomic system with four monochromatic classical microwave fields, where we indicate the respective Rabi frequencies and detunings between our levels by $\Omega_i$ and $\Delta_i$ for $i \in \{1,2,3,4\}$.}
	\label{fig1} 
\end{figure}

In the following we propose a setting for the quantum simulation of PEM interaction using Rydberg atoms, by engineering the interaction Hamiltonian in the ground energy subspace with four Rydberg levels interacting with four classical light fields as shown in Fig.~\ref{fig1}.
All classical light fields are monochromatic microwave laser, described by the overall field
\begin{equation}\label{eq1}
\begin{aligned}
\bm{E} &= \hat{\bm{\epsilon}}_1 E_1 \text{cos}(\omega_1 t + \phi_1) + \hat{\bm{\epsilon}}_2 E_2 \text{cos}(\omega_2 t + \phi_2) + \\ &\hat{\bm{\epsilon}}_3  E_3 \text{cos}(\omega_3 t + \phi_3) + \hat{\bm{\epsilon}}_4 E_4 \text{cos}(\omega_4 t + \phi_4)
\end{aligned}
\end{equation}
where $\hat{\bm{\epsilon}}_\alpha$ and  $\phi_\alpha$ are a unit polarization vector and relative phase of each field, respectively. The electric fields can then be decomposed into two exponential terms $\bm{E} = \bm{E}^{(+)} + \bm{E}^{(-)}$, where $\bm{E}^{(+)} =\sum_{j=1}^{4} \frac{1}{2} \hat{\bm{\epsilon}}_j E_j \text{e}^{-i(\omega_j t + \phi_j)}$ is the positive-rotating component and  $\bm{E}^{(-)} =\sum_{j=1}^{4} \frac{1}{2} \hat{\bm{\epsilon}}_j E_j \text{e}^{i(\omega_j t + \phi_j)}$ negative-rotating. 
Additionally, we assume that the wavelength of the field is much longer than the size of the atom. Hence, the spatial dependence of the field can be ignored over the size of the atom as per the dipole approximation.\\

Since the dipole operator $\mathbf{d} = -\text{e} \mathbf{r}_e$ is an odd parity operator, the diagonal elements of the operator vanish, and the (real) dipole matrix elements of each coupling pair are given by
\begin{equation}
\begin{aligned}
\mathbf{d} &= \bra{0}\mathbf{d}\ket{1} (\sigma_1 + \sigma_1^{\dagger})  + \bra{0}\mathbf{d}\ket{2} (\sigma_2 + \sigma_2^{\dagger}) + \\ &\bra{2}\mathbf{d}\ket{3} (\sigma_3 + \sigma_3^{\dagger}) +  \bra{0}\mathbf{d}\ket{3} (\sigma_4 + \sigma_4^{\dagger}),
\end{aligned}
\end{equation}
where $\sigma_1 = \ket{0}\hspace{-1mm}\bra{1}$, $\sigma_2 = \ket{1}\hspace{-1mm}\bra{2}$,  $\sigma_3 = \ket{2}\hspace{-1mm}\bra{3}$ and $\sigma_4 = \ket{0}\hspace{-1mm}\bra{3}$. Under free atomic evolution, the expectation values of $\sigma_{1,2,3,4}$ have unperturbed time dependence terms of $\text{e}^{-i\omega_{01}t}$, $\text{e}^{-i(\omega_{02}-\omega_{01})t}$, $\text{e}^{-i(\omega_{03}-\omega_{02})t}$, $\text{e}^{-i\omega_{03}t}$ which are all positive-rotating. These terms can be considered as $\mathbf{d}^{(+)}$. Similarly, the expectation values of $\sigma^{\dagger}_{1,2,3,4}$ have unperturbed time dependence of opposite sign, which therefore can be considered as $\mathbf{d}^{(-)}$, such that the dipole operator can be decomposed as $\mathbf{d} = \mathbf{d}^{(+)} + \mathbf{d}^{(-)}$. The interaction Hamiltonian caused by dipole interaction is given by $\hat{H}_{\text{int}} = -\mathbf{d} \cdot \bm{E}$. After applying the so-called rotating-wave approximation which focuses on slow dynamics rather than fast, thereby ignoring interaction terms $\mathbf{d}^{(+)} \cdot \bm{E}^{(+)}$ and $\mathbf{d}^{(-)} \cdot \bm{E}^{(-)}$, such that the interaction Hamiltonian becomes

\begin{equation}\label{eq3}
\begin{aligned}
\hat{H}_{\text{int}} &= -(\mathbf{d}^{(+)} \cdot \bm{E}^{(-)} + \mathbf{d}^{(-)} \cdot \bm{E}^{(+)}) \\
&= \frac{\Omega_1}{2}(\sigma_1 \text{e}^{i(\omega_1 t + \phi_1)} + \sigma^{\dagger}_1 \text{e}^{-i(\omega_1 t +\phi_1)}) \\
&+ \frac{\Omega_2}{2}(\sigma_2 \text{e}^{i(\omega_2 t + \phi_2)} + \sigma^{\dagger}_2 \text{e}^{-i(\omega_2 t +\phi_2)}) \\
&+ \frac{\Omega_3}{2}(\sigma_3 \text{e}^{i(\omega_3 t + \phi_3)} + \sigma^{\dagger}_3 \text{e}^{-i(\omega_3 t +\phi_3)}) \\
&+ \frac{\Omega_4}{2}(\sigma_4 \text{e}^{i(\omega_4 t + \phi_4)} + \sigma^{\dagger}_4 \text{e}^{-i(\omega_4 t +\phi_4)}), 
\end{aligned}
\end{equation}

where the Rabi frequencies for each coupling pair are $\Omega_1 = - \ket{0}\hat{\epsilon}_1 \cdot \mathbf{d}\ket{1} E_1$, $\Omega_2 = - \bra{1}\hat{\epsilon}_2 \cdot \mathbf{d}\ket{2} E_2$, $\Omega_3 = - \bra{0}\hat{\epsilon}_3 \cdot \mathbf{d}\ket{3} E_3$ and $\Omega_4 = - \bra{0}\hat{\epsilon}_4 \cdot \mathbf{d}\ket{3} E_4$. The free atomic Hamiltonian is defined as 

\begin{equation}\label{eq4}
\hat{H}_A = \omega_{01} \ket{1}\bra{1} + \omega_{02} \ket{2}\bra{2} + \omega_{03}\ket{3}\bra{3},
\end{equation}

where $\ket{0}$ denotes the zero energy ground state. The evolution of the system is calculated by solving Schr\"{o}dinger equation $i\partial_t \ket{\tilde{\psi}} = (\hat{H}_A + \hat{H}_{\text{int}}) \ket{\tilde{\psi}}$. The state $ \ket{\tilde{\psi}}$ is defined as the quantum state in the so-called rotating-frame. In this frame, we assume that the state $\ket{3}$ maintains the same velocity of its dynamics in the original frame, and the other states are sped up with different velocities such that $\ket{\tilde{\psi}} = \tilde{c_0} \ket{0} +  \tilde{c_1} \ket{1} +  \tilde{c_2} \ket{2} + c_3 \ket{3} $ where $\tilde{c_0} = \text{e}^{-i(\omega_1 + \omega_2 + \omega_3)t} c_0$, $\tilde{c_1} = \text{e}^{-i(\omega_2 + \omega_3)t} c_1$ and $\tilde{c_2} = \text{e}^{-i\omega_3 t} c_2$. 
The multiplied exponential factors are inserted to fasten the dynamics of states $\ket{0}$, $\ket{1}$ and $\ket{2}$. It can be shown that the resulting Hamiltonian under which the state $\ket{\tilde{\psi}}$ evolves is explicitly time-independent when the condition $\Delta_4 = \Delta_1 +\Delta_2+\Delta_3$ is satisfied, also known as the four-photon resonance condition. The time-independent Schr\"{o}dinger equation is given by
\begin{widetext}
\begin{equation}\label{eq5}
i\partial_t \begin{pmatrix}
\tilde{c_0} \\ \tilde{c_1} \\ \tilde{c_2} \\ c_3
\end{pmatrix} = \begin{pmatrix}
-(\Delta_1 + \Delta_2 + \Delta_3) & \frac{\Omega_1}{2}\text{e}^{i\phi_1} & 0 & \frac{\Omega_4}{2}\text{e}^{i\phi_4} \\ 
\frac{\Omega_1}{2}\text{e}^{-i\phi_1} &-(\Delta_2 + \Delta_3) & \frac{\Omega_2}{2}\text{e}^{i\phi_2} & 0 \\
0 & \frac{\Omega_2}{2}\text{e}^{-i\phi_2} & -\Delta_3 & \frac{\Omega_3}{2}\text{e}^{i\phi_3} \\
\frac{\Omega_4}{2}\text{e}^{-i\phi_4} & 0 & \frac{\Omega_3}{2}\text{e}^{-i\phi_3} & 0
\end{pmatrix} \begin{pmatrix}
\tilde{c_0} \\ \tilde{c_1} \\ \tilde{c_2} \\ c_3
\end{pmatrix},
\end{equation}
\end{widetext}
where for convenience the energy reference is adjusted such that the state $\ket{3}$ is the zero-energy level.
Since the ground subspace of the parafermion chain is three-dimensional, we can apply \textit{adiabatic elimination} to the system's equations of motion in Eq.~(\ref{eq5}), which yields the effective three-level Hamiltonian. Firstly, we have boosted by an overall energy $\Delta$, with $\Delta = (\Delta_1 + \Delta_2 +\Delta_3 +\Delta_4)/2$. Considering the equation of motion for $c_3$, when $\Delta \gg \Gamma$, or the natural decay happens very slow within the timescale of the internal atomic oscillation, this means $c_3$ carries the fast oscillation and it is supposed to be damped by coupling to the vacuum. so we can adiabatically eliminate $c_3$ by making the approximation that it damps to equilibrium instantaneously, i.e. $\partial_t c_3 = 0$. Therefore, we can obtain the substitution for $c_3$ by $\tilde{c}_0$, $\tilde{c}_1$, $\tilde{c}_2$ from the last equation of Eq.~(\ref{eq5}) such that $c_3 = - \left( \frac{\Omega_4}{2\Delta}\text{e}^{-i\phi_4}\tilde{c}_0 + \frac{\Omega_3}{2\Delta}\text{e}^{-i\phi_3}\tilde{c}_2 \right)$. Substituting this into the equation for $\tilde{c}_0$, $\tilde{c}_1$, $\tilde{c}_2$ in Eq.~(\ref{eq5}), we finally obtain the effective 3-level Hamiltonian where the state $\ket{3}$ is eliminated,
\begin{widetext}
\begin{equation}\label{eq9}
i\partial_t \begin{pmatrix}
\tilde{c_0} \\ \tilde{c_1} \\ \tilde{c_2} 
\end{pmatrix} = \begin{pmatrix}
-\left(\Delta_1 + \Delta_2 + \Delta_3 + \frac{\Omega^2_4}{4\Delta} \right) & \frac{\Omega_1}{2}\text{e}^{i\phi_1} & \frac{-\Omega_4 \Omega_3}{4\Delta}\text{e}^{i(\phi_4 - \phi_3)}\\ 
\frac{\Omega_1}{2}\text{e}^{-i\phi_1} &-(\Delta_2 + \Delta_3) & \frac{\Omega_2}{2}\text{e}^{i\phi_2} \\
\frac{-\Omega_4 \Omega_3}{4\Delta}\text{e}^{-i(\phi_4 - \phi_3)} & \frac{\Omega_2}{2}\text{e}^{-i\phi_2} & -\left(\Delta_3+\frac{\Omega^2_3}{4\Delta}\right) 
\end{pmatrix} \begin{pmatrix}
\tilde{c_0} \\ \tilde{c_1} \\ \tilde{c_2} 
\end{pmatrix}.
\end{equation}
\end{widetext}
Since states $\ket{0}$ and $\ket{2}$ initially interact directly with state $\ket{3}$, when $\ket{3}$ is adiabatically eliminated, the resulting effective Hamiltonian includes the effective Rabi coupling between state $\ket{0}$ and $\ket{2}$, which is $\Omega_R = \frac{-\Omega_4 \Omega_3}{2\Delta}\text{e}^{i(\phi_4 - \phi_3)}$. In addition, the energy terms are also shifted due to an AC Stark shift, amounting to $\Omega^2_4/(4\Delta)$ and $\Omega^2_3/(4\Delta)$ for state $\ket{0}$ and $\ket{2}$, respectively. 
In order to recover the parafermion interaction Hamiltonian given in Eq.~(\ref{eq:Hint}), the Rabi frequencies need to obey $\Omega_i = 2g \hspace{1mm} \forall i$, the detunings set to $\Delta_1=\Delta_3 = \Delta_4 = g$, $\Delta_2 = -g$, and relative phases of the laser fields following $\phi_1 = \phi_2 = -(\phi_4 - \phi_3) = \frac{2\pi}{3}$, where $g$ is a controllable factor depending on laser detuning. \\

By fixing the external field parameters according to these identities, we can implement the braiding of parafermions using this system, by realising a Berry phase evolution on states $\ket{2}$ and $\ket{3}$. This entails turning on solely the interaction between the latter states, and allowing for an adiabatic evolution in order for a geometric phase to accumulate, as described in Ref.~\cite{Pinto2009}, where it was shown that the adiabatic evolution of a two-level model in the presence of an external classical electric field yields the Berry phase
\begin{equation}
    \gamma_l = \frac{l}{2}\oint_0^T dt \frac{|D(t)|^2}{F_l(t)}\dot{\phi_3}(t),
    \label{eq:BerryPhase}
\end{equation}
where $l$ indicates the instantaneous eigenstates of the model, $D(t) = \bra{2}\mathbf{d}\ket{3}\cdot \hat{\epsilon}_3 E_3(t)$, $F_l(t) = \left(\frac{\omega_{23}}{2} \right)^2 + |D(t)|^2 - l \cdot \left(\frac{\omega_{23}}{2} \right)\sqrt{\left(\frac{\omega_{23}}{2} \right)^2 + |D(t)|^2}$, and $\phi_3(t)$ is the time-dependent relative phase of the coupling laser between state $\ket{2}$ and $\ket{3}$. While dynamical phases arise in this procedure, leading to unwanted dephasing in the time evolution, these can be closely monitored and compensated for. Indeed, to cancel out the dynamical phase accumulated in the adiabatic evolution, the other coupling channels are turned on, as shown in Fig.~\ref{fig1}. This allows apply adiabatic elimination to state $\ket{3}$, which yields the additional AC Stark shift to state $\ket{2}$ that could destructively compensate for the unwanted dephasing, leaving only the geometric phase. This corresponds to the Berry phase required to simulate the braiding of parafermion as derived in Eq.~(A6) in Ref.~\cite{Hutter_2016}. 

\section{\label{sec:Discussion} Discussion}

Motivated by the use of direct short-range interaction between Majorana quasi-particles to  achieve the $\frac{\pi}{8}$-phase gate on a topologically encoded qubit \cite{Bravyi_2005}, we investigated a similar prescription in the case of $\mathds{Z}_3$ parafermions. We studied the interaction between the edge modes of a parafermion chain using exact perturbation for low orders, in order to find its effect on the degenerate ground space of the system, which is used to encode one qutrit. We find that allowing the edge modes a degree of delocalisation facilitates the generation of a non-topological dynamical gate. In particular, the structure of this gate can directly be exploited to realise interesting non-Clifford operations such as the qutrit equivalent of the (qubit) $\frac{\pi}{8}$-phase gate (provided access to extra Clifford operations), as well as unitaries in higher levels of the Clifford hierarchy. This is a crucial complement to the topological operations since the addition of any non-Clifford gate to the Clifford group generates a set of unitaries that is dense in $SU(d^n)$ \cite{Campbell_2012}. As a sanity check, we additionally adopted recent results on the universality of one-qudit gates \cite{Sawicki_2017} in order to test our gate-set supplemented with the non-topological $\mathcal{U}$ with success. We illustrated this universality by sampling states obtained with Clifford operations (as provided by braiding parafermions \cite{Hutter_2016}), complemented with the dynamical non-Clifford gate $\mathcal{U}$, visualised using two parameter spaces which we defined to characterise qutrit states. While such representations served their purpose within our study, we refer the reader to the recent Ref.~\cite{Eltschka_2021} by Eltschka et. al. for a three-dimensional model of the qutrit state space that captures its prominent and essential geometric features. 

The non-Clifford gate accessible through PEM interaction is not topologically protected. However, fault-tolerance can be reinstated using stabilizer operations in a magic state distillation protocol \cite{Bravyi_2005, Campbell_2012, Anwar_2012}. Contextuality represents a critical resource for MSD and can be characterised by a state-independent measure $\mathcal{M}$ \cite{Howard_2014}. Using this result we showed that our gate-set provides non-contextual states under stabilizer measurements, which can be exploited in appropriate distillation routines. While we studied the $\mathds{Z}_3$ case for convenience, one could generalise the above analysis to $\mathds{Z}_d$ parafermion edge-mode interaction, particularly for prime odd $d$ as the emergent gates are of interest in the context of quantum information processing.

Finally, we showed how the two-parafermion (edge mode) interaction and braiding can be simulated using a four-level Rydberg system, offering an experimental setting for probing simulated topological qudits. Alternatively, it has been suggested to use ultra-cold molecules to realize qudits with different four vibrational levels as shown in \cite{Cornish2020} which benefits from its longer coherence time \cite{Cornish2021}. However, the strength of dipole-dipole interactions between molecules remains less significant than in the case of Rydberg atoms, making them more favourable to implement a qutrit. Finally, it is worth mentioning that quantum simulations with qutrits might yield a novel tool for studying quantum many-body physics such as quantum phase transitions and out-of-equilibrium phenomena. It is of interest that the Hamiltonian of many qutrit systems with dipole-dipole interactions can show quantum phase transitions or topological effects  which might yield  exotic results  in condensed matter physics. As an example, in recent work from Blok et. al. \cite{Blok2021} quantum information scrambling was witnessed on a superconducting processor which can provide insight into quantum chaos and black hole dynamics. 

\section{Acknowledgements}

A.B. acknowledges funding from the EPSRC Centre for Doctoral Training in Delivering Quantum Technologies at UCL, Grant No. EP/S021582/1L, D.E.B. was supported by the Quantera project
Quantum Codes Design and Architecture EPSRC grant EP/R043647/1, and J.K.P. from EPSRC Grant No. EP/R020612/1. T. and C. S. A. is supported by EPSRC grants EP/V030280/1, EP/V047302/1, EP/S015973/1, EP/R035482/1, EP/R002061/1, EP/P012000/1. 

\begin{appendix}

\vspace{2mm}

\section{\label{app:huh} The gate $\mathcal{U}$ in the Clifford hierarchy}

\hspace{-4mm}\textbf{Theorem 1.} \textit{Let $U = HVH^{\dagger}$ where $H\in \mathcal{C}^{(2)}$, $V \in \hspace{5mm} \mathcal{C}^{(n)}$ and $\mathcal{C}^{(k)}$ is $k$-th level of the Clifford hierarchy,} $ U \in \mathcal{C}^{(n)} $.

\vspace{4mm}

\hspace{-4mm}In order to prove this succinctly we define Lemma 1 below which follows from Eq.~(\ref{eq:hierarchy}) along with the observation that $H^3 = H^{\dagger}$. 

\vspace{4mm}

\hspace{-4mm}\textbf{Lemma 1.}\textit{ Let $U \in \mathcal{C}^{(k)}$ and $P \in \mathcal{C}^{(1)}$,  \begin{equation}
    (HUH^{\dagger})P(HU^{\dagger}H^{\dagger}) = H\tilde{U}H^{\dagger}
\end{equation} 
where $ \tilde{U} \in \mathcal{C}^{(k-1)}$.}

\vspace{4mm}

\textit{Proof of Theorem 1.} Let $V \in \mathcal{C}^{(n)}$ and $P$ be an arbitrary Pauli operator such that $VPV^{\dagger}\in \mathcal{C}^{(n-1)}$. We define the unitary operator $U = HVH^{\dagger}$. Using Lemma 1 the following ensues
\begin{equation}
    UPU^{\dagger} \sim H\beta_{n-1}H^{\dagger}
    \label{eq:A2}
\end{equation}
where $\beta_{i}$ is a unitary operator in the $i^{\text{th}}$ level of the Clifford hierarchy, i.e. $\beta_{i} \in \mathcal{C}^{(i)}$. The case for $n=2$ is straightforward as $VPV^{\dagger} \in \mathcal{C}^{(1)}$, from which we obtain
\begin{equation}
    (HVH^{\dagger})P(HV^{\dagger}H^{\dagger}) \in \mathcal{C}^{(1)}
\end{equation}
For general $n \geq 3$, one can start from Eq.~(\ref{eq:A2}) and again act on an arbitrary Pauli operator with $H\beta_{n-1}H^{\dagger}$ to find that 
\begin{equation}
    (H\beta_{n-1}H^{\dagger})P(H\beta_{n-1}^{\dagger}H^{\dagger}) \sim H\beta_{n-2}H^{\dagger}
\end{equation} 
where $P \in \mathcal{C}^{(1)}$. One can iterate this procedure $m $ times such that for $m = 0$ :
\begin{equation}
    (H\beta_{n}H^{\dagger})P(H\beta_{n}^{\dagger}H^{\dagger}) \sim H\beta_{n-1}H^{\dagger}
    \label{eq:A5}
\end{equation}
and subsequently for general $m$ :
\begin{equation}
    (H\beta_{n-m}H^{\dagger})P(H\beta_{n-m}^{\dagger}H^{\dagger}) \sim H\beta_{n-m-1}H^{\dagger}.
    \label{eq:A6}
\end{equation}
Finally when $m = n - 3$ :
\begin{equation}
    (H\beta_3H^{\dagger})P(H\beta_3^{\dagger}H^{\dagger}) \sim H\beta_2H^{\dagger} 
\end{equation} 
\vspace{-1mm}where one identifies $H\beta_2H^{\dagger} \in \mathcal{C}^{(2)}$. Using equation \ref{eq:A5} recursively, it is then straightforward to show that $H\beta_{n}H^{\dagger} \in \mathcal{C}^{(n)}$, and therefore $U \in \mathcal{C}^{(n)}$.

\end{appendix}

\bibliographystyle{unsrt} 
\bibliography{mybibliography.bib}

\vspace{10mm}
\pagebreak

\end{document}